\documentclass[epj]{svjour}
\usepackage{graphics}
\usepackage{amsfonts}
\usepackage{amsbsy}
\usepackage{mathrsfs}
\usepackage{amsmath}
\usepackage{amssymb}
\def\be{\begin{equation}}
\def\ee{\end{equation}}
\def\bea{\begin{eqnarray}}
\def\eea{\end{eqnarray}}

\def\ba{\begin{array}}
\def\ea{\end{array}}

\def\lsim{\mathrel{\rlap{
\lower4pt\hbox{\hskip-3pt$\sim$}}
\raise1pt\hbox{$<$}}}     
\def\gsim{\mathrel{\rlap{
\lower4pt\hbox{\hskip-3pt$\sim$}}
\raise1pt\hbox{$>$}}}            
\def\vec#1{\mbox{\boldmath $#1$}}

\begin{document}
\title{Transverse $\Lambda^0$ polarization in inclusive quasi-real photoproduction at the current fragmentation}
\author{I. Alikhanov\inst{1}\thanks{\email{ialspbu@mail.ru}} \and O. Grebenyuk\inst{2}
}                     
\institute{Saint Petersburg State University, Saint Petersburg
198904, Russia  \and Petersburg Nuclear Physics Institute,
Leningrad District, Gatchina 188350, Russia}
%
%
\abstract{It is shown that the recent HERMES data on the
transverse $\Lambda^0$ polarization in the inclusive quasi-real
photoproduction at $x_F>0$ can be accommodated by the strange
quark scattering model. Relations with the quark recombination
approach are discussed.
} 
\authorrunning{I. Alikhanov, O. Grebenyuk}
\titlerunning{Transverse $\Lambda^0$ polarization in quasi-real photoproduction}
\maketitle
%
%
\section{Introduction\label{introd}}
Polarization of $\Lambda^0$ hyperons has been under scrutiny
almost since the very moment of its discovery.  Investigations of
the phenomenon received especially great impetus in 1976 due to
the striking experimental results obtained at FERMILAB, where the
hyperons produced in $pN$ collisions at 300 GeV proton beam energy
were highly polarized \cite{fermilab}. The polarization was
transverse and negative, directed opposite to the unit vector
$\vec{n}\propto[{\vec p_{b}}\times{\vec p_\Lambda}]$, where $\vec
p_b$ and $\vec p_\Lambda$ are the beam and hyperon momenta,
respectively.

Only this direction is allowed by the parity conservation in
strong interactions provided the incident particles are
unpolarized (henceforth, under polarization we imply just
transverse one unless otherwise stated). The results turned out to
be in disagreement with predictions of perturbative QCD, no
polarization had been expected to play any significant role in
high energy processes as the helicity is conserved in the limit of
massless quarks.

The polarization has been also observed in other hadron-hadron
reactions at different kinematic regimes \cite{panagiotou}. Its
features qualitatively coincide in almost all the reactions, for
instance, being insensitive to the incident particle energy,
exhibiting roughly linear growth by magnitude with the hyperon
transverse momentum $p_T$ and being negative. The only known
exception is the $K^-p$ process, where the polarization sign has
been found to be positive \cite{kexp}.

Certainly, there have been proposed many models attempting to
account for the results, e.g.
\cite{anderson,degrand,swed,gago,cea,ryskin,fujita,soffer,preparata,troshin,qrm1,anselmino,dong},
however neither of them is able to describe the complete set of
the available measurements.

The $\Lambda^0$ wave function facilitates theoretical studies
allowing to describe the phenomenon in a reasonable way. The exact
SU(6) symmetry, requires the spin-flavor part of the wave function
to be combined of the $ud$ diquark in a singlet spin state and the
strange quark of spin 1/2, or formally
$|\Lambda\rangle_{1/2}=|ud\rangle_0|s\rangle_{1/2}$, where the
subscriptions denote the spin states. Therefore, the total spin of
$\Lambda^0$ is entirely given by the spin  of its valence $s$
quark. There is also an alternative look at the spin transfer in
fragmentation appeared after publication of the polarized deep
inelastic lepton-nucleon scattering (DIS) data of the EMC
Collaboration \cite{emc}. It suggests that the spin carried by the
valence quarks is only a part of the total nucleon spin, the rest
one being attributed, for example, to the orbital angular momenta
of the valence quarks and to the nucleon sea (sea quarks,
antiquarks and gluons). Which picture, SU(6) or DIS, is suitable
for the description of the process remains still an important
issue \cite{burkardt,jaffe,boros,kotzinian,liu,ellis}. The
$\Lambda^0$ hyperon can provide here a useful instrument for the
study of the spin effects in strong interactions.

We used the SU(6) approach throughout this paper. The choice was
dictated by a wish to keep the quark scattering model as it is,
encouraged by the SU(6) based calculations of the longitudinal
$\Lambda^0$ polarization in $e^+e^-$ annihilation at the $Z^0$
pole \cite{gustafson} and their successful experimental
verification \cite{aleph,opal}.

According to the empirical rules proposed by DeGrand and
Miettinen, the polarization sign depends on whether the $s$ quark
is accelerated (increases its energy) or decelerated (decreases
its energy) in the $\Lambda^0$ formation process \cite{degrand}.
To illustrate, there are no valence $s$ quarks in the initial
state of the $pp$ reaction so that they come from the quark sea to
form the final $\Lambda^0$. But the sea quarks predominantly
populate small $x$-states ($x$ is Bjorken variable) and
consequently increase their average energy coming in the valence
content of $\Lambda^0$. Here the polarization is negative.
Contrary, incident pseudoscalar kaons of the $K^-p$ reaction
already contain valence strange quarks mostly decelerated in the
hadronization because the average energy of the created
$\Lambda^0$-s is less than that of the $K^-$ beam. In this case
the sign is positive. Similar ideas were implemented in flux-tube
models with orbital angular momentum \cite{anderson}.

It was natural to wonder whether the polarization took place in
$\Lambda^0$ electro and/or photoproduction. Would one observe here
the same features with hadron-hadron reactions? These questions
have been investigated, for example, in experiments on high energy
$\gamma N$ scattering performed at CERN \cite{cern_gamma} and SLAC
\cite{slac_gamma} in the beginning of the eighties. However, their
statistical accuracy is indecisive and would hardly enable one to
conclude on the magnitude or on the sign of the polarization.

The transverse $\Lambda^0$ polarization has been also measured in
unpolarized $e^+e^-$ annihilations, for example, by TASSO
Collaboration  at 14, 22 and 34 GeV center-of-mass (cms) energies
\cite{tasso} and near the $Z^0$ pole by OPAL \cite{opal}. The
polarization observed in both experiments is consistent with zero.
Practically, this process can be a good place for deriving some
important information on the hadronization phase. In particular,
it can assist understanding in which extent final state
interactions contribute to the transverse polarization \cite{lu}.

In light of the scarce statistics for the $\Lambda^0$
photoproduction process, the HERMES experiments on the 27.6 GeV
positron beam scattering off the nucleon target acquires a
particular status providing a good opportunity for observation of
the polarization in electroproduction. The collaboration has
measured nonzero positive transverse polarization, when most of
the intermediate photons are quite close to the mass shell, i.e.
$Q^2=-(p_{ei}-p_{ef})^2\approx 0$ GeV$^2$ \cite{new}, where
$p_{ei}$ and $p_{ef}$ are the 4-momenta of the initial and
scattered electrons, respectively (quasi-real photoproduction).

Experimental properties of the polarization at HERMES turned out
to be very reminiscent of those in the $K^-p$ reaction
\cite{kexp}, which has been successfully described by a model
assuming the polarization to appear mostly via strange quark
scattering in a color field \cite{swed,gago}.

Thus, there are indications that mechanisms responsible for the
phenomenon in the $K^-p$ and $ep$ may be similar, at least, within
the covered kinematic region. These arguments inspired us to apply
the model to the $\Lambda^0$ quasi-real photoproduction data
obtained by HERMES.

Another goal of this paper is qualitatively to discuss some
relations between the calculations presented herein and the quark
recombination model (QRM) \cite{qrm1}.

\section{Quark scattering model for $\Lambda^0$\label{model}}
Electrons, when scattering off nuclei, have been known to be able
to acquire polarization. Analytically, it can be found within QED
by considering a process of Dirac pointlike particle scattering
off static Coulomb potential provided next-to-leading order
amplitudes are taken into account \cite{feshb,dalitz} (see also
Ref. \cite{dharma}). The corresponding formula reads
\be \vec{P}={2  \alpha_{em}    m p \over E^2}{ \sin ^3 {\theta/2}
\ln( \sin  {\theta/2}) \over [1-{p^2}/{E^2} \sin ^2 {\theta/2}]
\cos {\theta/2}}\vec{n}, \label{eq:sz1} \ee
where $E$, $p$, $m$ and $\theta$ are the energy, momentum by
magnitude, mass and scattering angle of the electron,
respectively, $\alpha_{em}$ is the fine structure constant,
$\vec{n}\propto[\vec{p}_i\times\vec{p}_f]$, $\vec{p}_i$ and
$\vec{p}_f$ are the vectors of the electron momenta in the initial
($i$) and final ($f$) states.

In Ref. \cite{swed}, Szwed proposed to explain the $\Lambda^0$
polarization as polarization of its valence strange quark in
scattering using Eq. (\ref{eq:sz1}), the $\Lambda^0$ wave function
favors such a consideration quite well. The idea is to perform the
following interchanges in Eq. (\ref{eq:sz1}): electron
$\leftrightarrow$ quark, $\alpha_{em}$ $\leftrightarrow$
$C\alpha_s$ (Coulomb potential $\leftrightarrow$ color field),
where $\alpha_s$ is the strong coupling and $C$ is the color
factor.

This approach has been applied to describe the polarization in the
$K^-p$ reaction and successfully reproduced its main features at
$2C\alpha_s=5.0$ and the $s$ quark mass $m_s=0.5$ GeV \cite{gago}.

Due to some peculiarities of the HERMES experiment, the
polarization was not measured in the traditional form of the
dependence on $x_F=2p_z/\sqrt{s}$ ($p_z$ is the longitude
component of the detected particle momentum, $\sqrt{s}$ is the
total cms energy). Therefore, we have expressed the model in terms
of the light cone variable $\zeta$ the available data depend on
\cite{new}. It is defined as
\begin{equation}
\zeta_{i(f)}=\frac{E_{i(f)}+p_{zi(f)}}{E_b+p_{zb}},
\label{eq:zeta}
\end{equation}
here the subscription $b$ denotes the beam. We note that $\zeta$
is invariant under Lorentz boosts being  useful in its
application.

According to recipes given in Ref. \cite{gago}, one should move to
a frame, where the magnitudes of the initial and final $s$ quark
momenta are the same (originally called $S$-frame). It is reached
by performing a Lorentz transformation along the proton momentum.
For this purpose, one can write
\be\label{eq:4vec_prod} (p_i\cdot p_f)=p^2(1-\cos\theta)+m_s^2,\ee
\be p_{Tf}=p_T=p\sin\theta, \label{eq:p_t} \ee
where $p_{i(f)}$ are the 4-momenta of the scattering quark,
$p_{Tf}$ is the transverse momentum of the scattered quark in the
center-of-mass frame of the $K^-p$ reaction, while
$p=\sqrt{E^2-m_s^2}$, $p_T$ and $\theta$ refer to the $S$-frame.

On the other hand, using Eq. (\ref{eq:zeta}) leads to
\begin{multline}
(p_i\cdot p_f)-m_s^2=\frac{m_s^2}{2}
\frac{(\zeta_i-\zeta_f)^2}{\zeta_i\zeta_f}\\
+\frac{1}{2}\left[p_{Ti}^2\frac{\zeta_f}{\zeta_i}
+p_{Tf}^2\frac{\zeta_i}{\zeta_f}\right]+(\vec{p_{T}}_i\cdot\vec{p_{T}}_f),
\label{eq:4vec_prod_zeta}
\end{multline}
where $(\vec{p_{T}}_i\cdot\vec{p_{T}}_f)$ is the ordinary scalar
product of the transverse momentum vectors.

Neglecting, as a first approximation, the incident quark
transverse momentum ($p_{Ti}$=0), after some algebra, one can
obtain from Eqs. (\ref{eq:4vec_prod})-(\ref{eq:4vec_prod_zeta})
that
\be\label{rel1} \cos {\theta \over 2} = { \xi  V_T^2  \over
(1-\xi)^2 + V_T^2  },\ee \be V={(1-\xi)^2 +V_T^2 \over
2\sqrt{\xi}\sqrt{(1-\xi)^2+(1-\xi)V_T^2}}, \label{rel2}\ee
with the variables $V_{(T)}$, and $\xi$ defined by
\bea\label{ksi_vt} V_{(T)}=\frac{p_{(T)}}{m_s}, \quad
\xi=\frac{\zeta_f}{\zeta_i}. \eea
By using relations (\ref{rel1}) and (\ref{rel2}), Eq.
(\ref{eq:sz1}) can be rewritten as
\begin{equation}
P(\xi, V_T)= - {2C\alpha_s V \over 1+V^2 \cos^2 {\theta/2} }  {
\sin ^3 {\theta/2}  \ln (\sin  {\theta/2})  \over \cos {\theta/2}
}. \label{basic formula}
\end{equation}

Note that the minus sign in Eq. (\ref{basic formula}) appeared to
satisfy the rule of DeGrand and Miettinen in the region of our
interest ($\xi<1$).

\section{Calculations and results\label{results}}

\begin{figure}[h]
\centering
\resizebox{0.4\textwidth}{!}{%
\includegraphics{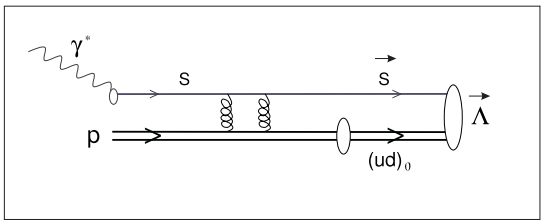}
}\caption{A schematic diagram of the $\Lambda^0$ polarization
process in quasi-real photoproduction. The $s$ quarks originated
from the photon scatters off the target color field, getting thus
polarized, and form the final hyperon recombining with a $(ud)_0$
diquarks from the proton target. An arrow over a letter indicates
the polarization.} \label{fig1}
\end{figure}

Our calculations concern the polarization in the region of
$0.25\lesssim\zeta_{\Lambda}\lesssim0.5$, which, according to Ref.
\cite{new}, corresponds to the events of $x_F>0$ in the cms frame
of the $\gamma^*p$ reaction (current fragmentation). The procedure
is now straightforward. In the model discussed above, one
substitutes the $K^-$ meson by the intermediate quasi-real photon
$\gamma^*$. Since the considered hyperons are produced in the
photon fragmentation region, we assumed that the $\Lambda^0$
kinematic is determined here in the main by the quarks originated
from the photon. The $(ud)_0$ diquarks are supposed to come mostly
from the proton target. The polarization process is schematically
shown in Fig. \ref{fig1}. Note that a similar diagram,
corresponding to the $s$ quark scattering off the scalar diquark
$(ud)_0$, appears also in the QRM \cite{qrm1} when one calculates
the polarization in the $K^-p$ reaction, however the interaction
is chosen to be scalar.

\begin{figure}[h]
\centering
\resizebox{0.4\textwidth}{!}{%
\includegraphics{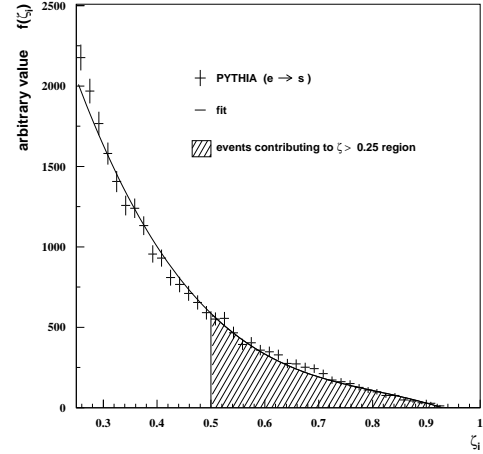}
} \caption{The $\zeta_i$ distribution of the quarks originated
from the positron beam according to the PYTHIA program (scattered
plot). The corresponding fit is presented by the solid line.
Events assumed to contribute to the region of $\zeta>0.25$ are
hatched.} \label{fig2}
\end{figure}

Within the present approach, one needs to know the $\zeta_i$
distribution of the incident $s$ quarks originated from the
quasi-real photons emitted by the positron beam. To find it, we
used the PYTHIA 6.2 program \cite{pythia62}, adopting thus the
positron-to-quark transition mechanisms implemented therein. The
obtained distribution is shown in Fig. \ref{fig2} (scattered plot)
together with its fit (solid line).

The final $s$ quark kinematic was determined according to the
following
\begin{eqnarray}
\zeta_f=\frac{m_s}{m_{\Lambda}} \zeta_{\Lambda}, \qquad
V_T=\frac{p_{T\Lambda}}{m_{\Lambda}}, \label{eq:qqm}
\end{eqnarray}
where $\zeta_{\Lambda}$ and $p_{T\Lambda}$ refer to the detected
$\Lambda^0$ hyperons. Let us omit in the sequel the index
$\Lambda$. The relations in Eq. (\ref{eq:qqm}) define the quark
momenta as fixed fractions of the corresponding final hyperon
momenta. In particular, the direction of the $s$ quark transverse
momentum is assumed to point in most cases in the direction of
$p_{T\Lambda}$ \cite{swed}. It is, of course, only an approximate
picture. One should take into account that the quarks are just
constituents of the hyperon but not free. In fact, a momentum
component of a quark inside a hadron is not fixed but has an
intrinsic distribution.

Note that there should be a threshold  for the $\Lambda^0$
production in the region of $\zeta>0.25$ initiated by the incoming
quarks due to the undetected hadron system always exists and
carries also away some part of the energy. In other words, not all
the $\zeta_i$ events will contribute to the $\zeta>0.25$ region.
Thus, in the calculations, we took into account only those quarks
with $0.5\leq\zeta_i\leq1$ (hatched area in Fig. \ref{fig2}).

As the experimental $\zeta$ dependence of the polarization is
available integrally over $p_{T}$, we additionally introduced
averaging over the transverse momentum of the detected
$\Lambda^0$-s. In this case, the polarization was determined as

\begin{equation}
P_{\zeta}=\int\hskip -0.3mm d\zeta_i\hskip 0.1mm dp_{T}\
h(p_{T})\hskip 0.3mm
P\left(\frac{\zeta}{\zeta_i},p_{T}\right)\hskip -0.6mm
f(\zeta_i).\label{eq:zeta_dep}
\end{equation}

Here, $h(p_T)$ and $f(\zeta_i)$ are the $p_T$ and $\zeta_i$
distribution functions of the detected $\Lambda^0$-s and the
incident quarks respectively, $P({\zeta}/{\zeta_i},p_{T})$ is the
polarization defined by Eq. (\ref{basic formula}).

For similar reasons, we determined the $p_T$ dependence of the
polarization as

\begin{equation}
P_{p_T}=\int\hskip -0.3mm d\zeta_i\hskip 0.1mm d\zeta\
g(\zeta)\hskip 0.3mm
P\left(\frac{\zeta}{\zeta_i},p_{T}\right)\hskip -0.6mm
f(\zeta_i),\label{eq:pt_dep}
\end{equation}

where $g(\zeta)$ is the $\zeta$ distribution function of the
detected hyperons.

Using Eqs. (\ref{eq:zeta_dep}) and (\ref{eq:pt_dep}), we carried
out the calculations. We used a typical $\zeta$ distribution of
the detected $\Lambda^0$-s ($0.25\leq \zeta \leq 0.5$). For
$h(p_T)$, we adopted that obtained by HERMES (0.2 GeV $\leq p_T
\leq 1.2$ GeV) \cite{ptdistr}.  Note that all the distributions
were prenormalized to unity. As the free parameter values, we have
taken $2C\alpha_s=5.0$ and $m_s=0.5$ GeV.

\begin{figure}[h]
\centering
\resizebox{0.4\textwidth}{!}{%
\includegraphics{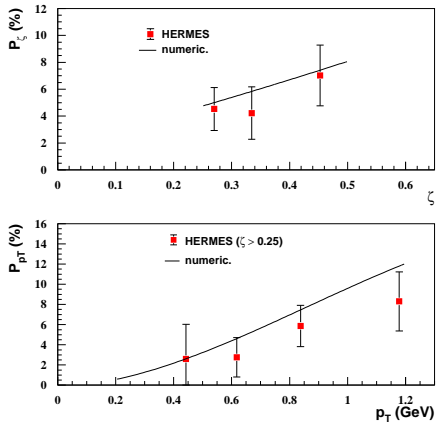}
} \caption{Numerical results (lines) in comparison with the HERMES
data (solid points). The $\zeta$ ($p_T$) dependence of the
$\Lambda^0$ polarization is presented in the upper (lower) panel.
The data are taken from Ref. \cite{new}.} \label{fig3}
\end{figure}

The numerical results in comparison with the HERMES data are shown
in Fig. \ref{fig3}. One can see that the experiment is reasonably
reproduced.

%
%
\section{Conclusion\label{conclusion}}

The results obtained herein should be regarded only as
qualitative. We neglected the transverse momentum of the incident
quarks ($p_{Ti}$), while, certainly, a strict consideration would
require taking it into account. However, for the goal, the present
work aimed at, such an approximation reflects the general
tendencies. It would be fair to expect a relatively narrow
$p_{Ti}$ distribution for the events contributing to the region of
$\zeta>0.25$.  To find the $\zeta_i$ distribution, we used the
PYTHIA program, which gives, in turn, rather qualitative than
quantitative predictions. We did not take the contributions from
the heavier resonances, such as $\Sigma^0$, $\Xi$ and $\Sigma^*$,
into account, their values are presumably considerable in
$\Lambda^0$ polarization \cite{gustafson,gatto,liang}. A
difficulty is also caused by the impossibility to derive the
running coupling constant $\alpha_s$ from the HERMES data.

In fact, the calculations by Eqs. (\ref{eq:zeta_dep}) and
(\ref{eq:pt_dep}) have been carried out similarly with the quark
recombination model \cite{qrm1}, the latter describes the
polarization in more quantitative manner. In the QRM, the central
point is the squared subprocess amplitude averaged over the
Bjorken variables in the initial and final states, their role, in
our case, were played by $\zeta_i$ and $\zeta$. A substantial
distinction between the QRM and the present quark scattering
approach (QSM) is the interaction.  For the QRM, it has been
assumed to be a scalar force, while the QSM calculations are based
on QCD. Thus it seems to be attractive alternatively to specify
the QRM interaction by the color one. Doing it could be regarded
as a further development of the present approach. It will include,
in particular, the transverse momentum of the incident quarks, the
structure functions of the projectile as well as the outgoing
hyperon instead of the approximations of Eq. (\ref{eq:qqm}), it
will also more explicate the underlying subprocesses. An
estimation of the contributions of the heavier resonances is in
progress now.


%
\listoffigures
\end{document}